\documentclass[aps,prl,twocolumn,amsmath,amssymb,superscriptaddress,showpacs,draft]{revtex4}
\usepackage{graphicx}
\input{epsf}

\newcommand{\I}{\hat{I}}
\newcommand{\el}{\hat{I}_{el}}
\newcommand{\EE}{\hat{I}_{ee}}
\newcommand{\pa}{\partial_{\bf p}}
\newcommand{\paa}{\partial^2_{ p_i,p_j}}
\begin{document}

\title{Ratchet transport of interacting particles}

\author{A.D.Chepelianskii}
\affiliation{\mbox{Laboratoire de Physique des Solides,
UMR CNRS 8502, B\^at. 510,
Universit\'e Paris-Sud, 91405 Orsay, France}}
\author{M.V.Entin}
\affiliation{\mbox{Institute of Semiconductor Physics,
Siberian Division of Russian Academy of Sciences,
Novosibirsk, 630090, Russia}}
\author{L.I.Magarill}
\affiliation{\mbox{Institute of Semiconductor Physics,
Siberian Division of Russian Academy of Sciences,
Novosibirsk, 630090, Russia}}
\author{D.L.Shepelyansky}
\affiliation{\mbox{Laboratoire de Physique Th\'eorique - IRSAMC, UPS \& CNRS,
Universit\'e de Toulouse, 31062 Toulouse, France}}

\date{August  21, 2008; Revised: September 17, 2008}


\begin{abstract}
We study analytically and numerically the
ratchet transport of interacting particles
induced by a monochromatic driving
in asymmetric two-dimensional structures.
The ratchet flow is preserved in
the limit of strong interactions
and can become even stronger compared to
the non-interacting case.
The developed kinetic theory gives a good
description of these two limiting regimes.
The numerical data show emergence of turbulence
in the ratchet flow under certain conditions.
\end{abstract}

\maketitle

\section{I Introduction}
For systems without spatial inversion symmetry the
appearance of directed flow of particles induced by a
time-periodic parameter variation with a zero-mean force
is now commonly known as the ratchet effect (see reviews
\cite{prost,reimann,hanggi2008}).
This phenomenon is ubiquitous in nature so that
such flows appear in a variety of systems
including asymmetric crystals
\cite{entin1978,belinicher} and semiconductor surfaces \cite{alperovich}
under light radiation, vortexes
in Josephson junction arrays \cite{mooij},
macroporous silicon membranes \cite{muller},
microfluidic channels \cite{ajdari} and others.
A significant increase of interest to ratchets
is related to the experimental progress in the investigation
of molecular transport in biological systems like proteins
characterized by asymmetry and non-equilibrium
\cite{prost,reimann,hanggi2008}. At the same time the
nanotechnology development allowed to fabricate artificial
asymmetric nanostuctures with the two-dimensional electron gas (2DEG)
where it has been shown that infrared or microwave radiation
creates a ratchet transport \cite{lorke,linke,song,ganichev}.
The theoretical studies predicted that the directionality of ratchet
flow in such systems can be controlled by the polarization of radiation
\cite{alik2005,dls2005,entin2006,alik2006,entin2007} 
that has been confirmed by recent experiments
with a semi-disk Galton board for 2DEG in AlGaAs/GaAs heterojunctions
\cite{portal2008}.

Till present the theoretical studies of ratchet transport have been done
mainly for non-interacting particles
\cite{prost,reimann,hanggi2008,entin1978,belinicher,alik2005,entin2007}.
However, in many systems the interactions between particles are of
primary importance like for example for microfluidic channels
\cite{ajdari}, 2DEG nanostructures
with strong electron-electron ({\it e-e}) 
interactions at a large $r_s$ parameter
\cite{kravchenko}, granulated materials \cite{meer} 
and one-dimensional Luttinger liquids \cite{braunecker}.
On a first glance it seems that a strong scattering between particles should
suppress the ratchet transport. But on the other hand the local conservation
of momentum of particles indicates
that even in presence of strong interactions
the ratchet flow still should exist.
The investigation of the properties of ratchet
transport for interacting particles 
in two dimensions is the main aim of this paper.
The theory developed is based on the kinetic approach
used in \cite{entin2007} extended to the case of strong interactions.
The theory is compared with the extensive numerical simulations
of ratchet transport of interacting particles in asymmetric structures.
The model description is given in Section II;
the analytical theory based on the kinetic equation is
developed in Section III; the numerical results are
presented in Section IV and the discussion is given in Section V.

\section{II Model description}
\begin{figure}
\centerline{\epsfxsize=7.0cm\epsffile{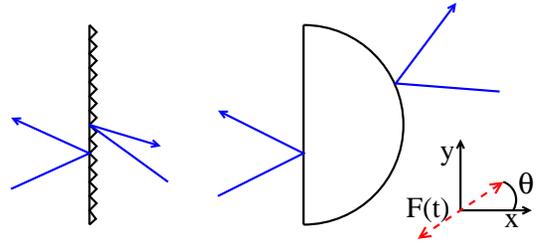}}
\vglue -0.4cm
\caption{(color online)Geometry of asymmetric scatterers
oriented in $(x,y)$-plane:
cuts with elastic (left) and diffusive (right) sides;
elastic semidisks; liner-polarized force $\mathbf{F}$
has angle $\theta$ in respect to $x$-axis.
}
\label{fig0}
\end{figure}

The interactions between particles are treated in the frame of the
mesoscopic multi-particle collision model (MMPCM) proposed by Kapral
(see e.g. \cite{kapral}). This method exactly preserves the total momentum
and energy of particles colliding inside each of $N_{cel}$ collision cells
on which the whole coordinate space with $N$ particles is divided.
In this method the collisions inside cells are modeled
by rotation of all particle
velocities in the moving center of mass frame of a given cell
on a random angle after a time $\tau_K$. To equilibrate the whole system
of interacting particles in presence of external monochromatic driving force
$\mathbf{F} \cos \omega t$ we use the Nos\`e-Hoover thermostat \cite{hoover}
which drives the system to the Boltzmann equilibrium
with a temperature $T=m{v_T}^2/2$
on a relaxation time $\tau_H$.
Such a combination of two methods for systems with interactions and
{\it ac-}driving has been already used in \cite{sync2007}.
As in \cite{entin2007}
the asymmetry appears due to asymmetric scatterers
having form of vertical cuts with diffusive (right) and elastic (left) sides
(cuts model) or of  elastic semi-disks of radius $r_d$
(semi-disks model) placed in a periodic square lattice of size $R \times R$.
The system orientation geometry and two types of scatterers
are shown in Fig.1 (see also \cite{entin2007} and Fig.~\ref{fig4} below).
In the cuts model it is assumed that the scattering on cuts takes
place instantaneously
at random moments of time which have a Poisson distribution with
time scale $\tau_c$. This corresponds to the case of
flashing cuts model (instantaneous appearance of cut at some moment of time)
which is slightly different from
the case of static cuts randomly distributed in the  plane
(both cases were discussed in \cite{entin2007}).
As in \cite{entin2007},
in absence of interactions an effective impurity scattering
is added with the scattering time $\tau_{im}$.
The monochromatic force is polarized as it is shown in Fig.~1
with  $\mathbf{F} = F (\cos \theta, \sin \theta)$.
Here, we present numerical results only for the 
semidisks model and the flashing cuts model,
which is rather convenient for numerical simulations, but in the
analytical treatment we also consider the static cuts model.
\begin{figure}
\centerline{\epsfxsize=7.0cm\epsffile{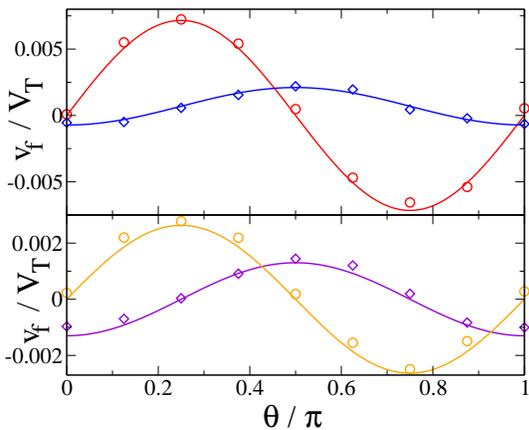}}
\vglue -0.4cm
\caption{(color online)Polarization dependence of the
average ratchet flow $\mathbf{v_f}$ in the flashing cuts model
with (top panel) and without (bottom panel) interactions;
diamonds and circles show numerical data  for $v_{f,x}$ and $v_{f,y}$
components, curves give the fits of data (see text).
The system parameters are: $N=10^4$, $N_{cel}=100 \times 100$
inside the periodic space domain $R \times R$ with
$v_T \tau_H/R=2.4$, $\tau_c/\tau_H=0.45$, $\tau_K/\tau_H=0.02$,
$\omega \tau_H=3$, $F v_T \tau_c/T=0.64$ for the top panel
and same parameters for the bottom panel but
$\tau_K/\tau_H =\infty$ and impurity scattering is added with
$\tau_{im}/\tau_H=0.5$; total integration time is $t/\tau_H \approx 10^3$.
}
\label{fig1}
\end{figure}

\begin{figure}
\centerline{\epsfxsize=8.0cm\epsffile{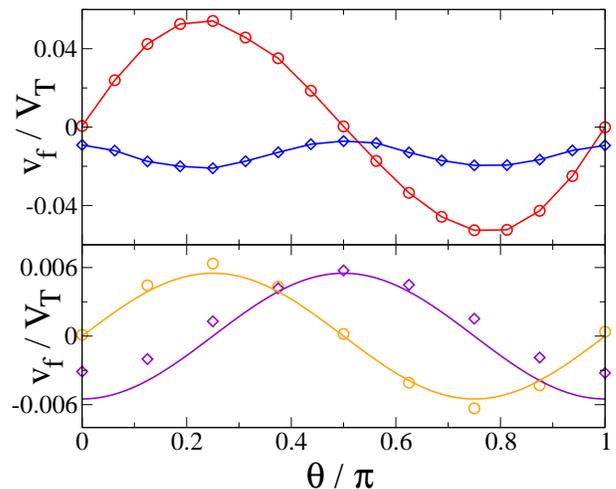}}
\vglue -0.4cm
\caption{(color online)
Same as in Fig.~\ref{fig1} for the semi-disk model with
$R/r_d=4$, $F r_d/T= 0.15$, $\omega \tau_H =1$,
effective $\tau_c/\tau_H \approx R^2/(2 r_d v_T \tau_H) =0.85$,
other parameters are as in Fig.~\ref{fig1}.
}
\label{fig2}
\end{figure}

The results of numerical simulations for
the polarization dependence of the ratchet flow in the flashing cuts model
are shown in Fig.~\ref{fig1}. In absence of interactions
the results are well described by the theory
\cite{entin2007} with the fit dependence
$\mathbf{v_f}/v_T = b (-\cos(2\theta), 2\sin(2\theta))/2$
where $b=0.0064 (F v_T \tau_c/T)^2 \approx 0.8 b_{th}$
and $b_{th}$ is the theory value (see Eqs.(9),(41) in \cite{entin2007}).
For interacting particles the fit gives the dependence
$\mathbf{v_f}/v_T = b_{int} (-a_1\cos^2\theta+a_2\sin^2\theta, \sin(2\theta))$
with $b_{int}/b = 2.7$ and $a_1 =0.10, a_2=0.29$. In presence of interactions
the ratchet flow appears even after polarization averaging.
The results for the semi-disks model are shown in Fig.~\ref{fig2}.
Without interactions the data are satisfactory described
by the theoretical dependence
$\mathbf{v_f}/v_T = b (-\cos(2\theta), \sin(2\theta))$
with the fitting value $b=0.24 (F r_d/T)^2 \approx 0.4 b_{th}$
and the theoretical value $b_{th}$ of \cite{entin2007}
(see Eq.(42) and discussion there).
In presence of interactions the polarization dependence of the flow
is qualitatively changed:  the component $v_y$ is enhanced by a
factor 8 and $v_x$ remains negative for all $\theta$
showing signature of 4th $\theta-$harmonic (Fig.~\ref{fig2},
top panel, curves are drown to adapt an eye).

\section{III Analytical theory}
The numerical simulations are based on the dynamical description of
motions of many interacting particles.  
To obtain an analytical description of the 
ratchet transport we use 
the kinetic equation approach valid for
systems with developed chaos and rapid decay of correlations. 
The validity of the kinetic equation requires
rare collisions with asymmetric scatterers (antidots)
and randomness of scattering events.
Under such conditions the kinetic equation can be
applied for comparative study with the numerical data
even if the numerical simulations are done
for a deterministic system with a periodic lattice of semi-disks 
of relatively large size.

The symmetry of the system  determines the ratchet flow which
mean velocity ${\bf v}_f$ is quadratic in the amplitude of the
{\it ac-}force ${\bf F}(t)= \mbox{Re}({\bf F} e^{-i\omega t})$.
Therefore, the flow velocity can be 
described  by the phenomenological expressions $$
 v_{f,x} =\alpha_{xxx}|F_x|^2+ \alpha_{xyy}|F_y|^2,~~~
 v_{f,y}= 2\mbox{Re}(\alpha_{yxy}F_xF_y^*)
.$$ The tensor components $\alpha_{xxx}$, $\alpha_{xyy}$ and
$\mbox{Re}(\alpha_{yxy})$ determine the response produced by a
linear-polarized monochromatic force ($\mbox{Im}{\bf F}=0$). 
In absence of interactions (see \cite{entin2007}), for the
linear polarization along $x$ or $y$ axes the mean flow is directed 
along $x$-axis; the current in $y$ direction appears for
tilted linear-polarized force. 

We also note that for the elliptically-polarized force
with $\mbox{Im}{\bf F}\neq 0$ there exists also a circular ratchet
effect determined by the product
of $\mbox{Im}(\alpha_{yyx})$ and $\mbox{Im}(F_xF_y^*)$ but we
will not consider this effect here.

 \subsection{Kinetic equation}
The kinetic equation in the momentum space $\mathbf{p}$ reads
\begin{equation}
\frac{\partial f}{\partial t}+{\bf F}(t)\frac{\partial f}{\partial \mathbf{p}} =\I (f) .
\label{eq1}
\end{equation}
where in the case of microwave field, ${\bf E}(t)$ is 
the electric field ${\bf E}(t)$
interacting with electron gas ${\bf F}(t)= e{\bf E}(t)$, $e$ is
the electron charge. The collision operator $\I=\el+\EE$ contains the
operator of elastic collisions (including impurities and 
scatterers (or antidots)) $\el$ and 
interparticle (electron-electron or {\it e-e}) collisions $\EE$.

The integral of elastic collisions with scatterers and static impurities reads as
\begin{equation}\label{}
\el(f_{\bf p})= \sum_{\bf p'}Q_{\bf pp'}f_{\bf p'}=
\sum_{\bf p'}[W({\bf p',p})f_{\bf p'}-W({\bf p,p'})f_{\bf p}],
\label{eq2}
\end{equation}
 where $Q_{\bf pp'}$ is the kernel of the operator
$\hat{I}_{el}$ and \ $W({\bf p',p})$ is the  probability of the
transition from ${\bf p'}$ to ${\bf p}$.

The interparticle collisions  operator ({\it e-e}) operator is
\begin{align}
\label{eq3}
& \EE (f)=\frac{2\pi }{S^2}
 \sum_{{\bf p}_1,{\bf p}',{\bf p}_1'} \delta_{{\bf p}+{\bf p}_1,{\bf p}'+{\bf p}_1'}  \\
& \times \delta(\epsilon_{{\bf p}}+\epsilon_{{\bf p}_1}-\epsilon_{{\bf p}'}-\epsilon_{{\bf p}_1'}) u^2_{{\bf p}-{\bf p}'} 
\nonumber \\
& \times \left\{f_{{\bf p}'}f_{{\bf p}_1'}(1-f_{{\bf p}})(1-f_{{\bf p}_1})
    -f_{{\bf p}}f_{{\bf p}_1}(1-f_{{\bf p}'})(1-f_{{\bf p}_1'})
    \right\} .
\nonumber
\end{align}
Here $S$ is the sample area, $u_{\bf k}$ is the Fourier transform
of {\it e-e}-interactions.

Interparticle collisions satisfy the conservation of the total
momentum of  gas. Due to the Galileo invariance the action of the
collision integral on the equilibrium distribution function with
shifted argument $\hat{I}_{ee}f^{(0)}_{\bf p+a}$ vanishes for any
${\bf a}$. Expanding by ${\bf a}$ we have:
\begin{eqnarray}
\label{eq4}
& \hat{I}_{ee}(f^{(0)}_{\bf p})=0,~~~\EE '({\bf a}\pa
f^{(0)})=0,\nonumber \\ 
& \EE''({\bf a}\pa f^{(0)}\ast{\bf
a}\partial_{\bf p'} f^{(0)})+\EE'(a_ia_j\paa f^{(0)})=0.
\end{eqnarray}
We use the following notations for the
first and the second variations around equilibrium: $\delta \EE
(f)=\EE'(\delta f)$ (linear operator), $\delta^2
\EE(f)=\EE''(\delta f\ast \delta f)$ (bi-linear operator, asterisk
denotes integration with two functions of different arguments).

The ratchet flow is generated by the anisotropy of collisions. 
This anisotropy is constructed
artificially due to asymmetric form of 
oriented scatterers. 
As theoretical models we considered cases of fixed
oriented anisotropic scatterers, namely cuts and semidisks. 
The model of static cuts is analytically
solvable \cite{entin2007}
but has a disadvantage since it leads to a divergence due
to electrons moving along the mirrors. Even if
this divergence can be regularized by an 
isotropic impurity scattering such a property is not 
very convenient.  Due to that it is useful to use a modified model
of flashing cuts which does not have such divergence.
In this model at any moment a particle
can meet a scatterer with a constant probability independent of
its velocity and direction of motion; after collision  the particle 
equi-probably scatters into any angle of the right semicircle  if it 
collides from the right semicircle and is mirror-reflected if it collides
from left semicircle (see Fig.1). Such a model of flashing cuts
gives a significant simplification for 
analytical and numerical studies.

The corresponding transition probability  in these models are
given by (see also \cite{entin2007}):
\begin{align}
 W({\bf p',p})=\frac{4\pi^2}{mS}w(\varphi',\varphi)\delta
(\varepsilon_{\bf p}-\varepsilon_{\bf p'}) ;
\label{eq5}
\end{align}
with
\begin{align} 
\label{cuts}   
& w(\varphi',\varphi)
=\frac{1}{\tau_c} \Big[\cos \varphi'~\theta(\cos \varphi')
   \delta(\varphi'+\varphi-\pi) \\
&    - \frac{1}{2}\cos \varphi' \cos \varphi~\theta(\cos \varphi)
    \theta(-\cos \varphi')\Big] \ \   (\mbox{static cuts}) ,
\nonumber
\end{align}

\begin{align}
     \label{sdiscs} 
& w(\varphi',\varphi) =\frac{1}{\tau_c}
     \Bigl[\cos \varphi'~\theta(\cos \varphi')\delta(\varphi'+\varphi-\pi)+ \nonumber \\
&    \frac{1}{4}|\sin{(\frac{\varphi-\varphi'}{2})}|[\theta(\varphi-\varphi')
    \theta(-\varphi- \varphi') \\
& + \theta(\varphi'-\varphi)\theta(\varphi+
    \varphi')]\Bigr] \ \ (\mbox{semi-disks}) ,
\nonumber
\end{align}

\begin{align}  
\label{flcuts}
 w(\varphi',\varphi) =\frac{1}{\tau_c} [\theta(\cos
\varphi')\delta(\varphi'+\varphi-\pi)+  
\nonumber \\
   \frac{1}{\pi}\theta(\cos \varphi)\theta(-\cos
    \varphi') ]  \ \ (\mbox{flashing cuts}) .
\end{align}
Here $\varphi$ is the polar angle of electron momentum
($-\pi<\varphi<\pi$), $\tau_c(\varepsilon)$ is the characteristic
scattering time on asymmetric scatterers, $\theta(x)$ is the
Heaviside function.

\subsection{Linear response} 
We consider the limit of high rate of interparticle scattering
exceeding the rate of elastic collisions. At the same time the
interactions preserve the total momentum and in isotropic media do
not affect the momentum relaxation. This is not the case for an
anisotropic medium where the interactions indirectly lead to the
momentum relaxation due to the conversion of the first angular
harmonic of the distribution function ${f\bf_p}$ to higher
harmonics produced by the anisotropic scattering. In particular,
it is generally excepted that in an isotropic medium with closed
Fermi surface the e-e scattering does not affect the conductivity.
Nevertheless, in the considered case of anisotropic medium e-e
collisions indirectly affect  the momentum relaxation rate. This
action is realized due to the conversion of the first angular
harmonics of the distribution function to higher harmonics
produced by the anisotropic scattering. As a result, the
conductivity becomes temperature dependent in the temperature
range when the e-e relaxation time is comparable with the elastic
relaxation time.

At first we consider the linear response to the electric field
using the expansion $f=f^{(0)}+f^{(1)}+f^{(2)}+...$
in small driving force $F$.
The linearized kinetic equation  can be written in the form
($f^{(1)}(t)= \mbox{Re}(f^{(1)}_\omega e^{i\omega t})$):
\begin{equation}\label{1}
    -i\omega f^{(1)}_\omega+{\bf F}_\omega \pa f^{(0)}=
\hat{I}^{(1)}(f^{(1)}_\omega),
\end{equation}
where the collision operator contains the elastic collisions with
anisotropic scatterers determined by $\hat{I}_{el}$ and  
interparticle or {\it e-e}
collisions determined by $\hat{I}'_{ee}$:
\begin{equation}\label{2}
   \hat{I}^{(1)}=\hat{I}_{el}+\hat{I}'_{ee}
\end{equation}
The formal solution of Eq.(\ref{1}) in the first order of
alternating force is
\begin{equation}\label{f1}
   f^{(1)}_\omega=( i\omega+\hat{I}^{(1)})^{-1} ({\bf
F}_\omega\pa) f^{(0)}.
\end{equation}
In the case of weak e-e interaction $\hat{I}'_{ee}$ can be canceled.
In the opposite limit of strong e-e scattering  
the formal parameter describing $\EE$ is large.
Having  in mind Eq.~(\ref{eq4}) we see that the inverse operator $(\omega+
\hat{I}^{(1)})^{-1}$  can be found by a projection on the subspace
of the Hilbert space of the basis functions $\psi_i=\frac{\partial
f^{(0)}}{\partial p_i}/||\frac{\partial f^{(0)}}{\partial p_i}||$
corresponding to  zero eigenvalue of the operator
$\hat{I}'_{ee}$. Thus the operator $\el$ is replaced by its
projection, while $\hat{I}'_{ee}$ can be canceled. The
resulting tensor of conductivity of {\it e-}charged particles with 
density $n_e$ reads
\begin{equation}\label{sig}
    \sigma_{ij}(\omega)=
\frac{e^2n_e}{m}\frac{\tau_i}{1-i\omega\tau_i}\delta_{ij},
\end{equation}
where  $\tau_i$ are relaxation times
of the first harmonics of the distribution function related with
the projected operator of elastic collisions:
\begin{eqnarray}\label{tau}
   \frac{ 1}{\tau_i}=-\sum_{\bf p,p'}\psi_i({\bf p})Q_{\bf pp'}\psi_i({\bf
    p'}) .
\end{eqnarray}
Here in $\tau_i$ index $i$ is axis index ($x$ or $y$). For the
considered systems from the relations (6)-(8) 
we have $\tau_i=\bar{\tau_c}/a_i$ and
\begin{eqnarray}\label{eqa}
    a_x=\frac{\pi}{8}+\frac{4}{\pi}, \ \ a_y=\frac{2}{3\pi} ~~(\mbox{for
  static cuts}), \nonumber \\
  a_x=\frac{2}{3}+\frac{8}{3\pi},\  \ a_y=\frac{2}{3}
 \nonumber ~~(\mbox{for
  semi-disks}), \nonumber\\
a_x=\frac{3}{2}+\frac{4}{\pi^2}, \ \ a_y=\frac{1}{2}~~ (\mbox{for
flashing  cuts}) .
\end{eqnarray}
The quantity $\bar{\tau_c}$ is determined by gas statistics:
$$\frac{1}{\bar{\tau_c}}=\frac{\int_0^\infty d\varepsilon
(f^{(0)'})^2(\varepsilon/\tau_c(\varepsilon))}{\int_0^\infty
d\varepsilon\varepsilon (f^{(0)'})^2} ,$$
where prime notes the derivative over the energy $\epsilon$.

In the case of static cuts or semi-disks $1/\tau_c(\varepsilon)
\propto \varepsilon^{s}$
with $s=1/2$. So one can write
$\bar{\tau_c}=\tau_c(\varepsilon_F)$ (strongly degenerate Fermi
case) and $\bar{\tau_c}=4\sqrt{2/\pi}\tau_c(T)/3$ (Boltzmann
case); $s=1/2$ for fixed obstacles and $s=0$ for flashing cuts (in
this case $\tau_c(\varepsilon)=const$).

The physical origin of Eqs. (\ref{sig}) and (\ref{tau}) is a very
quick relaxation of higher angular momenta harmonics as compared to 
the first harmonic relaxation. 
As a  result the conductivity has different values at low
temperature, when $\tau_{ee}\gg \tau_{el}$ and at high temperature
when $\tau_{ee}\ll \tau_{el}$. In both limits the conductivity
does not depend on e-e interaction, but has different values. In
the case of the Fermi distribution the conductivity changes from
low temperature value where $\tau_{ee}\gg \tau_{el}$ to high temperature
value where $\tau_{ee}\ll \tau_{el}$. We should emphasize that
the transition between these two values
is ruled by the ratio $\tau_{ee}/ \tau_{el}$ rather
than by the ratio of temperature $T$ to the Fermi energy $E_F$.
The transition temperature $T_0$ can be estimated 
by equating {\it e-e} relaxation time to the
relaxation time given by elastic scattering. In clean samples 
with high mobility the
transition corresponds to a rather low temperature $T_0\sim
E_F/\alpha~~\sqrt{\lambda_F/l_p}$, where $\alpha=(e^*)^2/\hbar
v_F$ is the dimensionless e-e interaction constant, $\lambda_F$
and $v_F$ are  the Fermi wavelength and velocity $l_p$ is the
elastic mean free path. For $E_F=0.01 eV$, $\lambda_F \approx 10 nm$,
$\alpha=0.5$, $l_p\sim 10^{-4}$ cm, $T_0\sim 10$K.

From Eq.(\ref{sig}) one can  write the expression for ratio of
static conductivities $\sigma_{yy}/\sigma_{xx}$:
\begin{equation}\label{anis}
    \sigma_{yy}/\sigma_{xx}=\tau_y/\tau_x = a_x/a_y.
    \end{equation}
In case of flashing cuts this ratio is equal to $3+8/\pi^2 \approx
3.81$ (see Eq.~(\ref{eqa})). For
such scatterers the problem of linear conductivity is solved
exactly also in the  limit of absence of e-e interaction
(see e.g. \cite{entin2007}). Using
Eq.(\ref{flcuts}), we find  $\sigma_{ii}=n_ee^2\tau_c b_i/m,
b_x=1/2, b_y=3/2$. Thus,  in this case
$\sigma_{yy}/\sigma_{xx}=3$. Hence, 
for example, for this flashing cuts  model 
the ratio $\sigma_{yy}/\sigma_{xx}=3$
is changed significantly when the temperature is changed from
$T< T_0$ to $T>T_0$.

\subsection{Quadratic response} 
The stationary ratchet flow appears in the second order
of {\it ac-}force $F$. In this case we can operate in a similar way
as before. The nonlinearity occurs due to the field term in the
kinetic equation and nonlinear e-e collision operator:
\begin{align}
\label{22}
 \partial_t f^{(2)}-(\el+\EE')(f^{(2)}) = \nonumber \\
-({\bf F}(t)\pa)f^{(1)}+\EE ''(f^{(1)}\ast
   f^{(1)}).
\end{align}
The projection of anisotropic elastic collision operator onto the
vector functions kills the third rank tensor needed for
photogalvanic current. So inclusion of anisotropy should be done a
bit more accurately. In short the stationary ratchet
current  is generated in a following way. 
The oscillating distribution function with
vector anisotropy is converted by nonlinear {\it e-e} interactions 
to the static second angular harmonics which in turn is partially
suppressed  by linear {\it e-e} interactions and then is
transformed to the
static vector anisotropy by anisotropic elastic collisions. 
The main contribution to the stationary flow reads
\begin{equation}\label{pge}
j_i=\frac{1}{S}\mbox{Re}\sum_{\bf p}v_i\I^{-1}\I^{(-)}\I^{-1}\EE^{(2)}(f_{-\omega}^{(1)}\ast f_\omega^{(1)}),
\end{equation}
The Eq.(\ref{pge}) has simplified form in accordance with the
smallness of the elastic antisymmetric operator $\I^{(-)}$ as
compared with the inelastic scattering ($\I^{(-)}$ obligatory
contains higher angular harmonics). 
The subsequent simplifications
include:  the substitution of $\EE'$ instead of  $\EE''$, 
according Eq. (\ref{eq4}); 
use of the
fact that inverse operators $\I^{-1}$ do not contain antisymmetric
operators; the cancellation of $\I^{-1}\EE^{(1)}$ acting on the
second angular harmonics; the replacement of the left operator
$\I^{-1}$ (taking into account summation with $v_i$) by the
inverse projected operator. As a result, we arrive at
\begin{equation}\label{100}
v_{f,i}=-\frac{1}{2}C\sum_{j,k}a_{jki}\tau_i\mbox{Re}(\tau_{\omega
j}\tau_{\omega k}^* F_{\omega j}F_{\omega k}^*).
\end{equation}
Here $\varepsilon=mv^2/2$ is the particle energy,
$a_{ijk}=<v_iv_j\I^{(-)}v_k>\tau_c/v^3$ is a numerical tensor,
characterizing the asymmetry of scatterers ($<...>$ stands for
average over angles in the momentum space), prime again means 
the derivative
over particle energy, $1/\tau_{\omega i}=-i\omega+1/\tau_i$. For
the specific cases of our models we obtain
\begin{eqnarray}\label{}
  a_{xxx}=\frac{1}{48}, \ \ a_{xyy}=-\frac{1}{16}~~ (\mbox{for
  static cuts}),\\
 a_{xxx}=-a_{xyy}=\frac{1}{12},
 ~~ (\mbox{for
  semi-disks}), \nonumber\\
a_{xxx}=\frac{1}{6\pi}, \ \ a_{xyy}=-\frac{1}{3\pi}, ~~(\mbox{for
flashing  cuts})\nonumber .
\end{eqnarray}
For $C$ we have:
$$ C=\frac{\int_0^\infty
d\varepsilon (f^{(0)'})^2(v^3/\tau_c)'}{\int_0^\infty
d\varepsilon\varepsilon (f^{(0)'})^2}.$$

In the case of static cuts or semi-disks
$C=(3/2+s)v_F^3/(\tau_c(\varepsilon_F)\varepsilon_F^2)$ (strongly
degenerate Fermi case) and $ C=2d_s(3/2+s)/(2^{1/2+s}m
\tau_c(T)\sqrt{mT})$ (Boltzmann case); $s=1/2, d_s=1$ for fixed
obstacles and $s=0,\ d_0=\sqrt{\pi}/2 $ for flashing cuts (in this
case $\tau_c(\varepsilon)=const$).

To compare with results of numerical calculations it is convenient
to write expressions for ratchet velocity components. For the linear
polarization of monochromatic force we obtain
\begin{align}
    \label{Vf} 
& v_{fx}/v_T= -B (F v_T \bar{\tau_c}/T)^2 a_{xxx} \times 
    \nonumber \\
& [\cos^2 \theta/(a_x^3(1+\omega^2\tau_x^2)) - 
\sin^2 \theta/(a_x a_y^2 (1+\omega^2\tau_y^2))] 
    \nonumber \\
& v_{fy}/v_T= -B (F v_T \bar{\tau_c}/T)^2 a_{xyy}  \times 
    \nonumber \\
& \sin(2\theta) (1+\omega^2 \tau_x \tau_y)/(a_x a_y^2 
(1+\omega^2\tau_x^2) (1+\omega^2\tau_y^2)) \; ,
\end{align}
where $B=C T^2 \bar{\tau_c}/2 v_T^3$ 
 and we remind that $\tau_i=\bar{\tau_c}/a_i$.
For the flashing cuts model we have
$\tau_c=\bar{\tau_c}$,
$a_{xxx}=1/6\pi$, $a_{xyy}=-1/3\pi$, $a_x=3/2+4/\pi^2$, $a_y=1/2$,
$C=2\sqrt{\pi}/(2m\tau_c \sqrt{2mT})$, $\tau_c=const$
 and for the semi-disks model
$a_{xxx}=-a_{xyy}=1/12$, $a_x=2/3+8/3\pi$, $a_y=2/3$,
$C=2/(m\bar{\tau_c}\sqrt{mT})$, $\bar{\tau_c} \propto T^{-1/2}$.
Here we give the results for the Boltzmann distribution $f^{(0)}$,
but similar calculations work for other $f^{(0)}$, e.g. for the
Fermi-Dirac distribution. 
We also give a simplified
derivation of the ratchet flow in the Appendix.
It is based on the local equilibrium distribution
and give the same results as Eqs.~(\ref{Vf}).

The opposite limit in absence {\it e-e} interactions was analytically
studied  for the cases  of static cuts \cite{entin2006} (exactly)
and approximately, for weak anisotropy,  for static cuts or
semi-disks \cite{entin2007}.  It is
important to emphasize that in both limits of weak and strong 
{\it e-e} interactions the current does not contain  the strength of 
interactions.  The transition between the regimes occurs when the
interparticle scattering rate becomes comparable with 
the rate elastic scattering on anti-dots and impurities. 

\section{IV Numerical results}

\begin{figure}
\centerline{\epsfxsize=8.0cm\epsffile{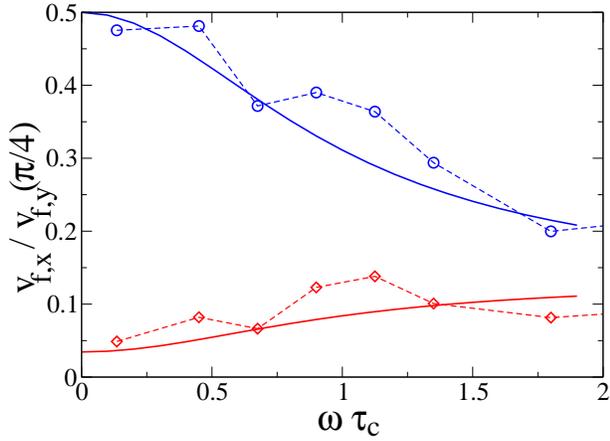}} \vglue -0.4cm
\caption{(Color online) Comparison between theory (\ref{Vf}) 
(full curves, no adjustable parameters) and numerical data for
interacting particles in the flashing cuts model (symbols); circles are for
$v_{f,x}$ and $\theta=\pi/2$ (here $v_{f,x} > 0$), 
diamonds are for $|v_{fx}|$ and
$\theta=0$ (here $v_{f,x}<0$ and we use the absolute value 
of $v_{f,x}$ in the ratio $v_{f,x}/v_{f,y}(\pi/4)$);
 $v_{f,y}$ is taken at $\theta=\pi/4$; other parameters
are as in Fig.~\ref{fig1}, top panel.} 
\label{fig3}
\end{figure}

\begin{figure}[h]
\centerline{\epsfxsize=5.5cm\epsffile{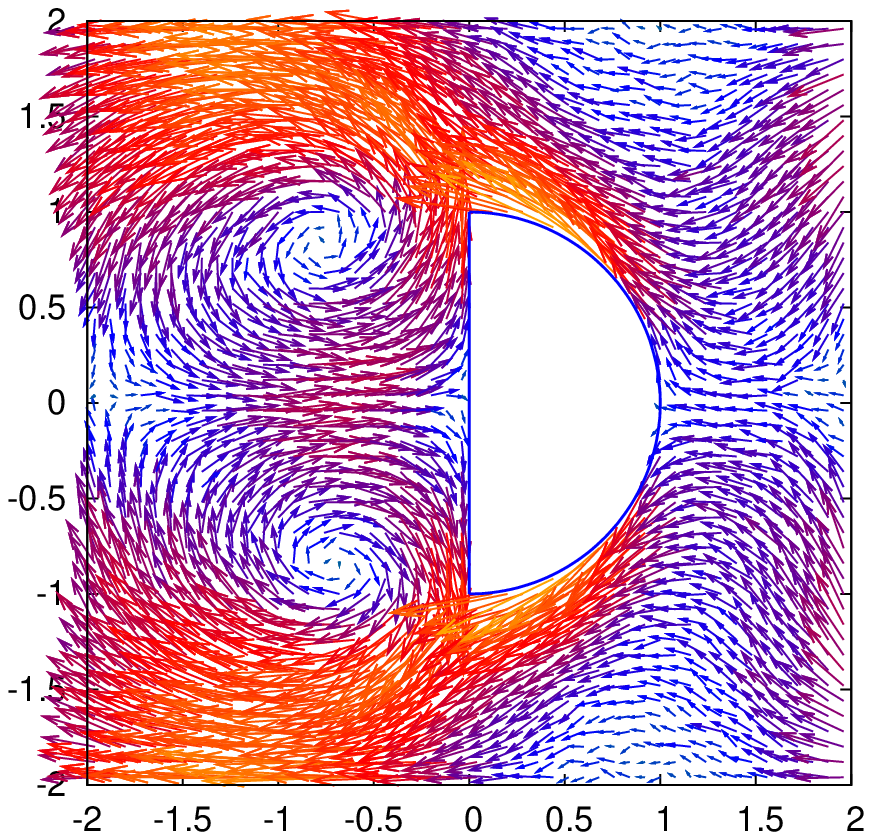}
\hglue -0.8cm
\hfill\epsfxsize=5.5cm\epsffile{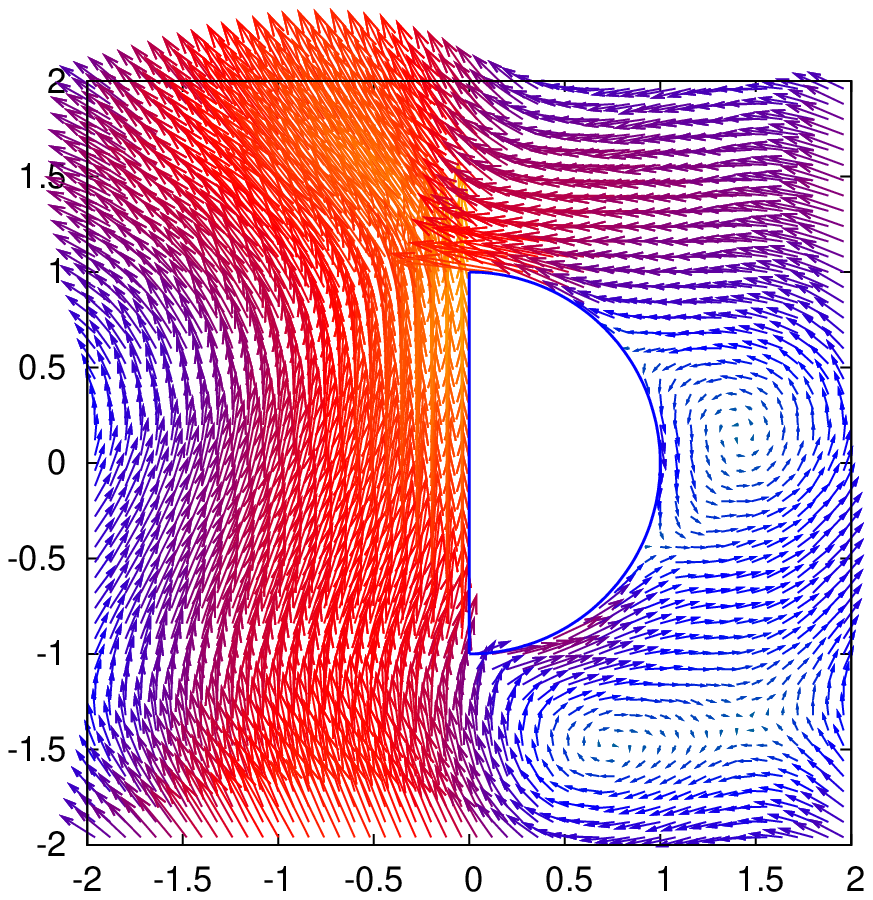}}
\centerline{\epsfxsize=5.5cm\epsffile{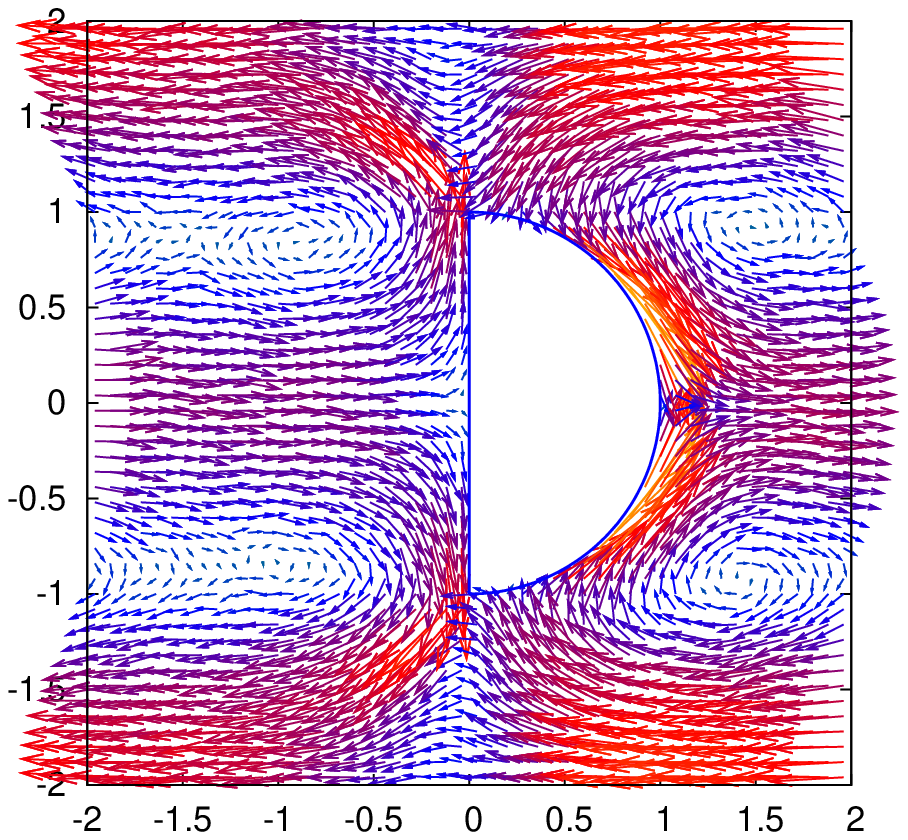}
\hglue -0.8cm
\hfill\epsfxsize=5.5cm\epsffile{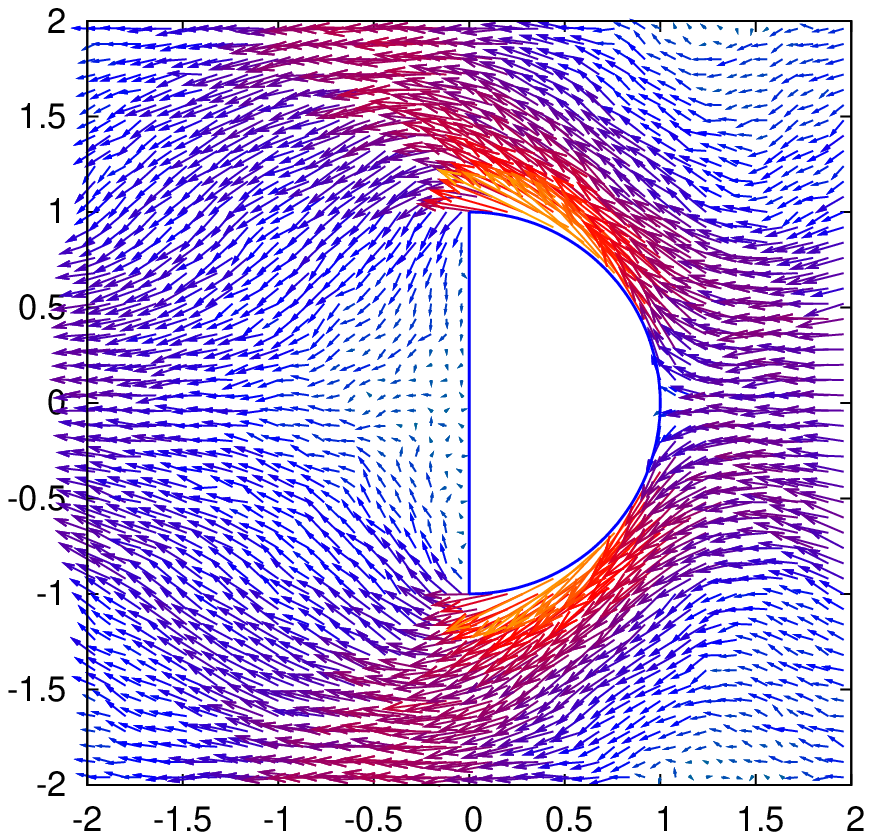}}
\vglue -0.30cm
\caption{(Color online) Map of local averaged velocities
in ($x$/$R$, $y$/$R$) plane of the semi-disks model
for parameters of Fig.~\ref{fig2} (top panel)
at $\theta=0$ (top left); $\theta=\pi/4$
(top right); $\theta=\pi/2$ (bottom left);
$\theta=0$ and 25 times increased interaction time
compared to other panels
($\tau_K/\tau_H=0.5$, $\tau_K v_T/r_d=4.7$ point in Fig.~\ref{fig5}).
The velocities are shown by arrows which size is proportional
to the velocity amplitude, which is also indicated by color
(from yellow/gray for large  to blue/black for small amplitudes).
}
\label{fig4}
\end{figure}

For the flashing cuts model the theory (\ref{Vf}) gives a good description
of numerical data (see Fig.~\ref{fig3}). For the semi-disks model
the agreement between the theory and numerical simulations
(Fig.~\ref{fig2}, top) is less accurate, e.g. 4th
$\theta$-harmonic for $v_{f,x}$ is absent in (\ref{Vf}). To
understand the origins of this difference we present the map of
local flow velocities at various polarizations $\theta$ in
Fig.~\ref{fig4}. For $\theta=0$ the results clearly show the
appearance of turbulent flow with two vertexes behind the
semi-disk. When the interaction scattering time $\tau_K$ is
increased by a factor 25 the interaction and turbulence
practically disappear and the average local flow becomes laminar
(see Fig.~\ref{fig4} top left and bottom right panels). At the
same time even with strong interactions the flow has much more
laminar structure for $\theta=\pi/4$ (Fig.~\ref{fig4}, top right
panel) when the absolute value of the total ratchet velocity has
its maximal value (see Fig.~\ref{fig2} top panel). Thus the
ratchet flow of interacting particles has certain similarities
with a hydrodynamic flow of the Navier-Stocks equation around
semi-disk body \cite{landau}. However, for $\theta=\pi/2$ the
ratchet flow is composed from two alternative flows moving in
opposite directions at the cell boundaries and the semi-disk
center (Fig.~\ref{fig4}, bottom left panel), such a rather flow is
different from hydrodynamic flows with fixed velocity far from the
body. For a qualitative description of the turbulent flow we may
argue that the turbulence leads to a difference of pressures on
different sides of the scatterer producing different resistances
for different flow directions. This generates the ratchet flow for
the {\it ac-}force driving. In general the kinetic description is
applicable when the interaction scattering length is large
compared to the scatterer size, e.g. $v_T \tau_K > r_d$  for
semi-disks. At small values of $\tau_K$ this condition is broken
(Figs.~\ref{fig2},\ref{fig4}) and we have transition to the
hydrodynamic like regime where the theory (\ref{Vf}) gives only
approximate description. For the flashing cuts model the kinetic
description remains always valid since the size of scatterer is
zero.
\begin{figure}
\centerline{\epsfxsize=8.0cm\epsffile{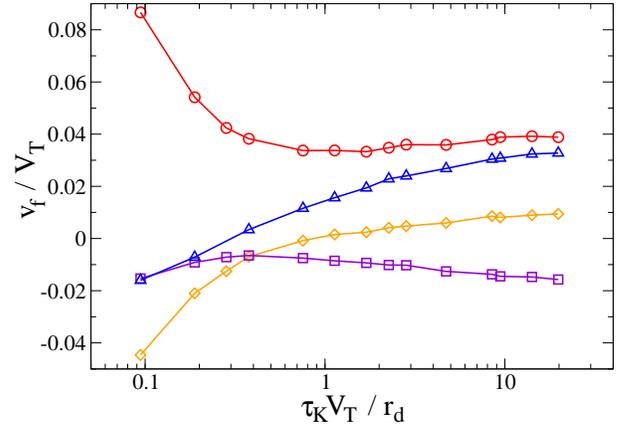}}
\vglue -0.2cm
\caption{(color online)
Dependence of the ratchet
velocity $v_f$ on the Kapral interaction scattering time
$\tau_K$ in the semi-disk model, numerical data are shown
by symbols: $v_{f,y}/v_T$ (red circles) and
$v_{f,x}/v_T$ (yellow diamonds) for $\theta=\pi/4$;
$v_{f,x}/v_T$ (violet squares) for $\theta=0$;
$v_{f,x}/v_T$ (blue triangles) for $\theta=\pi/2$;
other parameters are as in Fig.~\ref{fig2}, top panel,
curves are drown to adapt an eye.
}
\label{fig5}
\end{figure}

The dependence of the ratchet velocity on the interaction
scattering time $\tau_K$ is shown in Fig.~\ref{fig5}. The increase
of interactions (small $\tau_K$) can change the sign of the flow
in $x$-direction that is in a qualitative agreement with the
theory (\ref{Vf}). For weak interactions the flows are opposite in
$x$ for polarization $\theta=0$ and $\theta=\pi/2$ while at strong
interactions they are collinear. Thus in 2DEG in AlGaAs/GaAs
heterojunctions where interactions are relatively weak ($r_s \sim
1$) the flows are opposite for two polarizations in agreement with
the experiment \cite{portal2008}, but for other materials with
stronger interactions (e.g. SiGe with $r_s \approx 6$) the flows
may become collinear. We also note that at strong interactions the
rescaled ratchet characteristics are not sensible to the
temperature variation that indicates that we have an effective
liquid flow with temperature independent viscosity.

\section{V Discussion}

In conclusion, our extensive numerical simulations
show that even  in the regime of strong interactions
between particles a stationary ratchet flow is
generated by monochromatic driving in the asymmetric
periodic arrays. The obtained result are well
described by the analytical theory based on
the kinetic equation for strongly interacting particles.
It is interesting to note that for asymmetric arrays
the tensor of conductivity becomes temperature dependent 
due to interplay of interactions and relaxation of 
high momentum harmonics (see Eq.~(15) and discussion there). 
It would be interesting to investigate the effects of interactions 
on ratchet transport in experiments similar to those 
of \cite{portal2008}.

This research is supported in part by the ANR projects
MICONANO and NANOTERRA (France) and RFBR NN 08-02-00152-a,
08-02-00506-a (Russian Federation).

\renewcommand{\theequation}{A-\arabic{equation}}
  \setcounter{equation}{0}  
\section{APPENDIX}
Here, on the example of the flashing cuts model we give a more
simple and heuristic derivation of the ratchet flow
compared to the exact kinetic equation approach (\ref{eq1}).
In the regime of very strong interactions
we can assume that the ensemble of particles
is in a local equilibrium state
and hence the distribution function  can be written as
\begin{align}
f(\mathbf{v}, t) &= f_0(\mathbf{v} - \mathbf{v_0}(t))
\label{eqa1}
\end{align}
where $\mathbf{v_0}(t)$ is the instantaneous velocity of the center  of mass.
We put here the particle mass $m=1$.
It is assumed that the interactions give rapid relaxation
to the local equilibrium distribution
$ f_0(\mathbf{v} - \mathbf{v_0}(t))$.
The matrix of conductivity can be determined from
the momentum balance between acceleration
created by a small applied static force $\mathbf{F}$
and momentum loss on the asymmetric cut scatterer:
\begin{align}
&\frac{d \mathbf{p}}{d t} =
\mathbf{F} - \frac{1}{\tau_c}
\left( < \mathbf{V_c}(\mathbf{v})
f_0( \mathbf{v} - \mathbf{v_0} ) > - \mathbf{v_0}
\right) = 0 \, ,
\end{align}
where $ \mathbf{V_c}(\mathbf{v}) $ is the vector of average velocity
after a scattering with the incident velocity $\mathbf{v}$.
It is expressed via the scattering probability
$W({\bf v',v})$ (see the main text above) as
$\mathbf{V_c}(\mathbf{v}) =\int d \mathbf{v'} W({\bf v',v}) \mathbf{v'}$.
This gives the relation
\begin{align}
&< \mathbf{V_c}(\mathbf{v}) f_0( \mathbf{v} - \mathbf{v_0} ) >  = \\
    \nonumber
&\int d \mathbf{v} \left(
\begin{array}{c}
\frac{2 |\mathbf{v}| \theta(-v_x) }{\pi} -  v_x \; \theta(v_x) \\
v_y \theta(v_x) \end{array}
\right) f_0( \mathbf{v} - \mathbf{v_0} )
\end{align}
where $\theta(v)$ is the Heaviside function.
In the linear response regime we can expand in $\mathbf{v_0} $
that gives $f_0( \mathbf{v} - \mathbf{v_0} ) = f_0( \mathbf{v} ) + f_0( \mathbf{v} ) \frac{ \mathbf{v} \mathbf{v_0} }{T} + ...$ .
After integrating over the Maxwell distribution $f_0( \mathbf{v})$
we obtain
\begin{align}
&< \mathbf{V_c}(\mathbf{v}) f_0( \mathbf{v} - \mathbf{v_0} ) >  = \\
    \nonumber
&\int d \mathbf{v} \left(
\begin{array}{c}
\frac{2 |\mathbf{v}| \theta(-v_x) }{\pi} -  v_x \; \theta(v_x) \\
v_y \theta(v_x) \end{array}
\right) f_0( \mathbf{v} ) \frac{ \mathbf{v} \mathbf{v_0} }{T}
\\
    \nonumber
&= \left(
\begin{array}{c}
-\frac{8 + \pi^2}{2 \pi^2} v_{0,x} \\
\frac{1}{2} v_{0,y} \end{array}
\right)
\end{align}
where $\mathbf{v_0}$ is the velocity of the stationary flow.
Then the moment balance gives
\begin{align}
\left( \begin{array}{c}
v_{0,x}  \\
v_{0,y}
\end{array} \right)
 = \tau_c \left( \begin{array}{c}
\frac{2 \pi^2}{8 + 3 \pi^2} F_{x}  \\
2 F_{y}
\end{array} \right)
\end{align}
and therefore $\sigma_{yy}/\sigma_{xx}=\tau_y/\tau_x=a_x/a_y=3+8/\pi^2$
that is in agreement with the kinetic theory result given in the main text.
It is interesting to note that
for the noninteracting particles we have $\sigma_{yy}/\sigma_{xx}=3$
(see \cite{entin2007}).

To compute the ratchet flow we should expand the local velocity in
Eq.~(\ref{eqa1})
up to the second order in the driving force $F$:
$ \mathbf{v_0}(t) = {\hat \tau}_i \mathbf{F}(t) + \mathbf{v_f}$ ,
$v_f = O(F^2) $ where $\tau_i$ are the above values $\tau_x, \tau_y$
given by the linear response; we note that
second frequency harmonics $e^{\pm i 2 \omega t}$
are eliminated by the time averaging.
Then the time averaged distribution function is
\begin{align}
f(\mathbf{v}) &= <f(\mathbf{v}, t)>_t = \\
    \nonumber
& f_0(\mathbf{v}) + f_0(\mathbf{v}) \frac{ \mathbf{v} \mathbf{v_F} }{ T } +
\frac{ (\mathbf{v} {\hat \tau}_i \mathbf{F})^2 - T ({\hat \tau}_i \mathbf{F})^2  }{4T^2} f_0(\mathbf{v})
\label{eqa2}
\end{align}
where $f_0(\mathbf{v}) = \frac{1}{Z} \exp( - \frac{\mathbf{v}^2}{2 T} )$
is the Maxwell distribution and $\mathbf{v_F}$ is
the average ratchet flow velocity.
Again, the time averaged momentum balance equation reads
\begin{align}
\left( < \mathbf{V_c}(\mathbf{v})
f( \mathbf{v}) > - \mathbf{v_f}
\right) = 0 \, .
\end{align}
Using Eq.~(\ref{eqa2}) we obtain from the second term
the contribution $< \mathbf{V_c}(\mathbf{v}) f_0(\mathbf{v}) \frac{ \mathbf{v} \mathbf{v_F} }{ T } > = ((1-a_x) v_{fx}, (1-a_y) v_{fy})$
which is similar to the linear response term.
The integration of the third Gaussian term
gives the additional contribution
$(F^2/8\sqrt{2\pi T})
[(-\tau_x^2\cos^2 \theta + \tau_y^2 \sin^2 \theta), 2\tau_x \tau_y  \sin(2\theta) ]$. Finally we obtain
\begin{align}
\left(
\begin{array}{c}
v_{fx}  \\
v_{fy}\end{array}
\right) =
\frac{F^2}{8\tau_c \sqrt{2\pi T}}\left(
\begin{array}{c}
 -\tau_x^3 \cos^2 \theta + \tau_x \tau_y^2 \sin^2 \theta   \\
2 \tau_{x} \tau^2_y \sin(2 \theta) \end{array}
\right)
\end{align}
For $\omega \rightarrow 0$ these expressions are in agreement with
Eqs.~(\ref{Vf}) obtained by the kinetic equation theory.


\begin{thebibliography}{99}
\bibitem{prost} F. J\"ulicher, A. Ajdari, and J. Prost, Rev. Mod. Phys.
         {\bf 69}, 1269 (1997).
\bibitem{reimann} P.~Reimann, Phys. Rep. {\bf 361}, 57 (2002).
\bibitem{hanggi2008} P.~H\"anggi, and F.~Marchesoni, arXiv:0807.1283[cond-mat]
         Rev. Mod. Phys. to appear (2008).
\bibitem{entin1978} E.~M.~Baskin, L.~I.~Magarill, M.~V.~Entin,
        Sov. Phys.-Solid State {\bf 20}, 1403 (1978)
        [Fiz. Tver. Tela {\bf 20}, 2432 (1978)].
\bibitem{belinicher} V.~I.~Belinicher, B.~I.~Sturman, Sov. Phys. Usp.
        {\bf 23}, 199 (1980) [Usp. Fiz. Nauk {\bf 130}, 415 (1980)].
\bibitem{alperovich} V.L.Alperovich, V.~I.~Belinicher, V.~N.~Novikov, and
        A.~S.~Terekhov, JETP Lett. {\bf 31}, 546 (1980).
\bibitem{mooij} J.B.~Majer, J.~Peguiron, M.~Grifoni, M.~Tusveld,
        and J.E.~Mooij, Phys. Rev. Lett. {\bf 90}, 056802 (2003);
        A. V. Ustinov, C. Coqui, A. Kemp, Y. Zolotaryuk,
        and M. Salerno, Phys. Rev. Lett. {\bf 93}, 087001 (2004).
\bibitem{muller} S.~Matthias and F.~M\"uller, Nature {\bf 424}, 53 (2003).
\bibitem{ajdari} V.~Studer, A.~Pepin, Y.~Chen, and A.~Ajdari,
        Analyst {\bf 129}, 944 (2004).
\bibitem{lorke} A.~Lorke, S.~Wimmer, B.~Jager, J.P.~Kotthaus, W.~Wegscheider,
        and M.~Bichler, Physica B {\bf 249-251}, 312 (1998).
\bibitem{linke} H.~Linke, T.E.~Humphrey, A.~L\"ofgren, A.O.~Sushkov,
        R.~Newbury, R.P.~Taylor, and P.~Omling,
        Science {\bf 286}, 2314 (1999).
\bibitem{song} A.M.~Song, P.~Omling, L.~Samuelson, W.~Seifert, I.~Shorubalko,
        and H.~Zirath, Appl. Phys. Lett. {\bf 79}, 1357 (2001).
\bibitem{ganichev} P. Olbrich, E.L. Ivchenko, T. Feil, R. Ravash,
        S.D. Danilov, J. Allerdings, D. Weiss, S.D. Ganichev,
        preprint arXiv:0808.1983[cond-mat] (2008).
\bibitem{alik2005} A.D.Chepelianskii, and D.L.Shepelyansky,
         Phys. Rev. B {\bf 71}, 052508 (2005).
\bibitem{dls2005} G.~Cristadoro, and D.L.Shepelyansky,
         Phys. Rev. E {\bf 71}, 036111 (2005).
\bibitem{entin2006} M.~V.~Entin, L.~I.~Magarill,  
        Phys. Rev. B {\bf 73}, 205206 (2006).
\bibitem{alik2006} A.D.Chepelianskii, Eur. Phys. J. B {\bf 52}, 389 (2006).
\bibitem{entin2007} A.~D.~Chepelianskii, M.~V.~Entin, L.~I.~Magarill and
        D.~L.~Shepelyansky, Eur. Phys. J. B {\bf 56}, 323 (2007).
\bibitem{portal2008} S.~Sassine, Y.~Krupko, J.-C.~Portal, 
      Z.~D.~Kvon, R.~Murali,
         K.P.Martin, G.Hill, and A.D.Wieck,
        Phys. Rev. B {\bf 78}, 045431 (2008).
\bibitem{kravchenko} E.~Abrahams, S.~V.~Kravchenko, and M.~P.~Sarachik,
        Rev. Mod. Phys. {\bf 73}, 251 (2001).
\bibitem{meer} D. van der Meer, P.~Reimann, K. van der Weele, and D.~Lohse,
        Phys. Rev. Lett. {\bf 92}, 184301 (2004).
\bibitem{braunecker} D.~E.~Feldman, S.~Scheidl, and V.~M.~Vinokur
        Phys. Rev. Lett. {\bf 94}, 186809 (2005);
        B. Braunecker, D. E. Feldman, and J. B. Marston,
        Phys. Rev. B {\bf 72}, 125311 (2005).
\bibitem{kapral} A.~Malevanets,  and R.~Kapral,
        Lect. Notes Phys. (Springer) {\bf 640}, 116 (2004).
\bibitem{hoover}  W.~G.~Hoover, {\it Time reversibility, computer simulation,
        and chaos}, World Scientific, Singapore (1999).
\bibitem{sync2007} A.~D.~Chepelianskii, A.~S.~Pikovsky and D.~L.~Shepelyansky,
        Eur. Phys. J. B {\bf 60}, 225 (2007).
\bibitem{landau} L.~D.~Landau and E.~M.~Lifshitz, {\it Hydrodynamics},
        Nauka, Moskow (1986),

\end{thebibliography}
\end{document}